\documentclass[aps,prl,reprint,superscriptaddress]{revtex4-2}
\usepackage[utf8]{inputenc}
\usepackage[T1]{fontenc}

\usepackage{amsmath,amssymb}
\usepackage[version=4]{mhchem}
\usepackage{graphicx}
\usepackage{dcolumn}
\usepackage{bm}
\usepackage{siunitx}
\usepackage{booktabs}          
\graphicspath{{figures/}}
\usepackage{gensymb}
\usepackage{color}
\usepackage{xcolor}

\usepackage{multirow}
\usepackage[normalem]{ulem}
\usepackage{makecell} 
\usepackage{tabularx}
\usepackage{xr}
\usepackage{threeparttable}

\makeatletter
\long\def\figurecaption#1{%
  \setlength{\parindent}{0pt}%
  \setbox\@tempboxa\hbox{#1}%
  \ifdim \wd\@tempboxa > \hsize
    \justifying
    #1\par
  \else
    \centering
    \box\@tempboxa
  \fi
}
\makeatother

\newcommand{\CrMnSn}{\ce{(Cr_{x}Mn_{1-x})_{3}Sn}}
\newcommand{\MnSnplusd}{\ce{Mn_{3$\pm$\delta}Sn}}

\begin{document}

\title{Magnetic phase evolution and Hall effect modulation in Cr-doped stoichiometric \ce{Mn3Sn}}

\author{Padraig Maderson}
\affiliation{Debye Institute for Nanomaterials Science, Utrecht University, 3508 TA Utrecht, the Netherlands}
\author{Zhiyuan Cheng}
\affiliation{Leiden Institute of Physics, Leiden University, Niels Bohrweg 2, 2333 CA Leiden, the Netherlands}
\author{Iris Serra} 
\affiliation{Leiden Institute of Physics, Leiden University, Niels Bohrweg 2, 2333 CA Leiden, the Netherlands}
\author{Romain Sibille}
\affiliation{PSI Center for Neutron and Muon Sciences, CH-2532 Villigen-PSI, Switzerland}
\author{Jiacheng Zhang} 
\affiliation{State Key Laboratory of Quantum Functional Materials, School of Physical Science and Technology, ShanghaiTech University, Shanghai 201210, China}
\author{Yangfeng Guo} 
\affiliation{State Key Laboratory of Quantum Functional Materials, School of Physical Science and Technology, ShanghaiTech University, Shanghai 201210, China}
\affiliation{ShanghaiTech Laboratory for Topological Physics, ShanghaiTech University, Shanghai 201210, China}
\author{Navid Qureshi}
\affiliation{Institut Laue-Langevin, 71, avenue de Martyrs, 38021 Grenoble, France}
\author{Semonti Bhattacharyya}
\affiliation{Leiden Institute of Physics, Leiden University, Niels Bohrweg 2, 2333 CA Leiden, the Netherlands}
\author{Henrik Jacobsen}
\email{henrik.jacobsen@ess.eu}
\affiliation{Nanoscience Center, Niels Bohr Institute, University of Copenhagen, 2100 Copenhagen, Denmark}
\affiliation{Data Management and Scientific Computing, Asmussens Allé 305, 2800 Kongens Lyngby, Denmark}
\author{Machteld E. Kamminga}
\email{m.e.kamminga@uu.nl}
\affiliation{Debye Institute for Nanomaterials Science, Utrecht University, 3508 TA Utrecht, the Netherlands}

\date{\today}

\begin{abstract}
\noindent The antiferromagnet \ce{Mn3Sn} is a well-established candidate for spintronics, showing a very large anomalous Hall effect at room temperature. The origin of this effect stems from its topological nature, which in turn originates from the underlying 120° inverse-triangular antiferromagnetic structure. Here, we present Cr-doping as a direct tool to tune the anomalous Hall effect, via distortions in the magnetic structure. By synthesising a series of \CrMnSn\ with varying $x$ and perfect 3:1 stoichiometry of the transition metals to Sn, and performing SQUID magnetometry and electrical transport measurements, we find that Cr-doped \ce{Mn3Sn} exhibits a rich phase diagram. Small doping levels of $x \approx 0.04$ completely suppress the incommensurate magnetic phase of the undoped parent compound, and greatly reduce the coercive field for switching the anomalous Hall effect by an order of magnitude, to a record value for this family of compounds. Higher doping levels, on the other hand, give rise to ferromagnetic behaviour. Supported by single-crystal neutron diffraction, we show that \ce{(Cr_{0.08}Mn_{0.92})_{3}Sn} has a magnetic phase transition near 170 K with a reduction in symmetry from $Cmc'm'$ to $C2'/c'$, allowing for the magnetometry-observed ferromagnetic canting of the spins out of the $ab$-plane. Moreover, this concentration yields strong temperature-dependent transport behaviour. Our results highlight the complex coupling between doping and anomalous Hall response, allowing for vastly tunable spintronic properties of this topological Weyl semimetal through Cr doping, while maintaining stoichiometry.  
\end{abstract}

\maketitle

\section{Introduction}

\begin{figure}[h]
\centering
\includegraphics[width=0.45\textwidth]{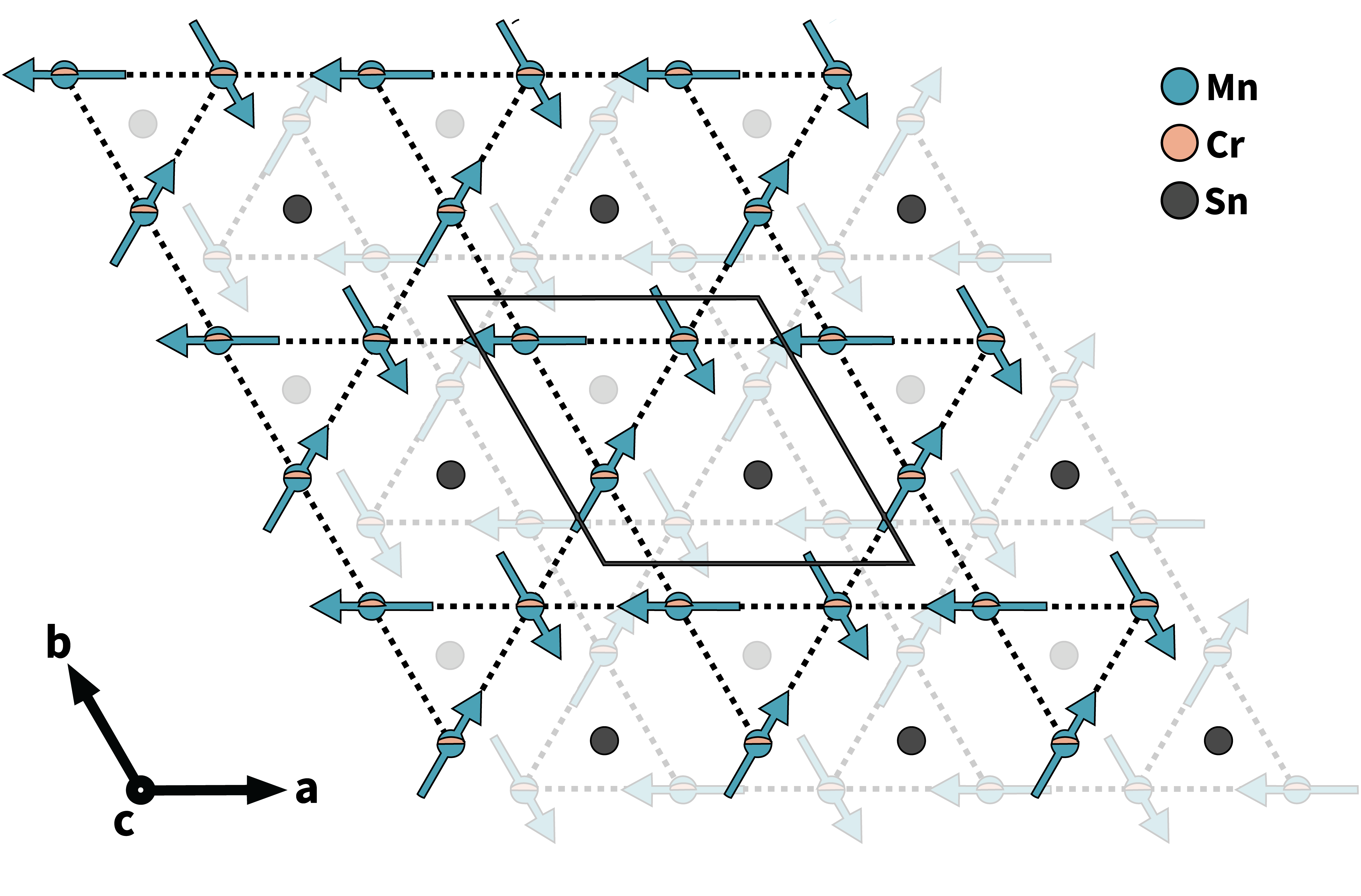}
\caption{Crystal structure of the 120° inverse triangular antiferromagnetic (IT-AFM) structure of \CrMnSn. Cr substitutes on the Mn sites and is depicted as small orange slivers on the blue Mn atoms. The unit cell is traced by black lines and the kagom\'e lattice indicated by dotted lines. }
\label{fig:kagome_structure}
\end{figure}

\noindent The Weyl semimetal \ce{Mn3Sn}, a member of the non-collinear antiferromagnets (AFM) \ce{Mn3X} ($X$ = Ga, Ge, Sn), has attracted significant attention by displaying a large $|\rho_H|= 6 \mu\Omega\cdot\text{cm}$ jump in anomalous Hall effect (AHE) at room temperature \cite{nakatsuji2015large,kiyohara2016giant,nayak2016large}. This is unique transport behaviour to occur spontaneously in an AFM with near zero net magnetisation \cite{chen2014anomalous}, and provides a handle to switch Hall voltage with small magnetic fields \cite{nakatsuji2015large}. In addition to its potential use for AFM magnetic memory devices due to lesser stray fields and faster switching than its ferromagnetic (FM) counterparts, \ce{Mn3Sn} has further collected a portfolio of other novel functionalities, including but not limited to ultrafast switching via light pulses \cite{ogawa2025ultrafast} and magnetic spin oscillator behaviour \cite{Takeuchi2021}. These functionalities stem from topological properties and broken inversion symmetry induced by the chiral inverse-triangular antiferromagnetic (IT-AFM) structure \cite{Kbler2014, Kuroda2017, chen2014anomalous} (Figure \ref{fig:kagome_structure}) making \ce{Mn3Sn} a prime candidate for integration into future spintronic devices.

Efforts to tune the AHE have been made using applied pressure \cite{Singh2020} and chemical (co)doping with \ce{Fe} \cite{Yano2024}, \ce{Cr} \cite{Yano2024,Liu2020,Zhang2026_Cr} or \ce{Co} \cite{Zhang2026_Co}. Both studies on Cr-doping \ce{Mn3Sn} found that doping led to the stabilisation of the IT-AFM phase at low temperatures \cite{Yano2024,Liu2020,Zhang2026_Cr}. However, they both report a (Cr+Mn):Sn ratio of around 3.1:1 or 3.1:0.9, and the seminal paper by Nakatsuji \textit{et al.} already reports that a small excess of Mn could stabilise the IT-AFM phase, resulting in the observation of the AHE down to low temperatures \cite{Nakatsuji2015}. Therefore, it is crucial to study the doping in stoichiometric \ce{Mn3Sn} in order to elucidate its true effect on the AHE. 

In this work, we perform a systematic doping study of Bi-flux grown \CrMnSn\ single crystals with 3:1 stoichiometry and \textit{x} ranging from 0 to 0.15. Using SQUID magnetometry, we show and thereby confirm the stabilisation of the IT-AFM phase with low Cr concentrations (\textit{x} = 0.04 - 0.05) and an abrupt first order change to FM behaviour at higher doping concentration (\textit{x} = 0.08 - 0.15). For \ce{(Cr_{0.08}Mn_{0.92})_{3}Sn}, our neutron diffraction experiments (performed in zero field) indicate that $C2'/c'$ symmetry fits equally as well as $Cmc'm'$, while allowing more degrees of freedom to explain the FM modes both parallel and perpendicular to the $ab$-plane that we see in our magnetometry. Most notably, the anomalous Hall transport behaviour shows strong doping dependence of the coercive field ($B_c$) and strong temperature dependence of $B_c$ for \textit{x} = 0.08 specifically. These findings provide concrete insights into tuning the $B_c$ needed to switch the polarity of the AHE and this will facilitate the implementation of \CrMnSn\ and its topological properties in future technologies.

As mentioned, the magnetic and anomalous transport properties of \ce{Mn3Sn} depend heavily on the exact Mn:Sn ratio, which tends to deviate from a perfect 3:1 ratio, resulting in structural formula \MnSnplusd\ \cite{Zimmer1972,kren1975study,Cable1993}. Czochralski grown crystals, i.e. single crystals pulled from a melt using a rotating seed, generally have slight excess Mn ($\delta=+0.02$) and show a suppression of the incommensurate phase, maintaining the IT-AFM phase at low temperatures \cite{Nakatsuji2015}. The Bridgman method, i.e. single crystals grown in a crucible translated through a temperature gradient, produces varying Mn excess and deficiencies ($\delta=+0.09$ to $-0.017$) \cite{Park2018,Cable1993}. With Sn-flux growth, the crystals have significantly varying magnetic behaviour as observed between the work of Sung \textit{et al.} \cite{Sung2018} and Yan \textit{et al.} \cite{Yan2019} (though true compositions are unspecified). Due to the high temperatures needed for the aforementioned growth methods, substitution of Mn on the Sn sites is thermodynamically favourable, resulting in an excess of Mn and associated magnetic effects \cite{Park2025}. The Bi-flux method, however, requires lower temperatures of max 1150\celsius \ facilitating the ideal \ce{Mn3Sn} stoichiometry \cite{cederholm2026ground,Park2025}. Therefore, we have chosen the Bi-flux method to provide the synthetic basis for this research, yielding stoichiometric \ce{Mn3Sn} crystals that eliminate substantial excess/deficiency as contributing factors. From this vantage point, we map the magnetic and topological features across the rich phase diagram of Cr-doped \ce{Mn3Sn}.

\section{Experimental Methods}
\noindent Single crystals of \CrMnSn\ ($x$ = 0, 0.04, 0.05, 0.08, 0.13, 0.15) were grown using a \ce{Bi}-flux method \cite{Park2025}. Stoichiometric mixtures of \ce{Mn} powder (Macklin, 99.9\% purity), \ce{Cr} powder (Macklin, 99.5\% purity), \ce{Sn} granules (Macklin, 99.999\% purity) and \ce{Bi} granules (Macklin, 99.999\% purity) were combined in a molar ratio of (Mn+Cr):Sn:Bi of 3:1:3, with Mn and Cr together totalling to 3. The mixture was sealed in an alumina crucible under a partial argon atmosphere within a quartz ampoule and heated to 1150\celsius \ ramping over 15 h and dwelling for 10 h. The temperature was then slowly cooled to 700\celsius \ at a rate of 5\celsius \ $\text{h}^{-1}$ and annealed for an additional 10 h. Then the temperature was further reduced to 300\celsius \ at a rate of 2\celsius \ $\text{h}^{-1}$ and held for 50 h. Finally, the crystals were isolated from the residual Bi flux via centrifugation.

Elemental analysis was performed using Inductively Coupled Plasma Optical Emission Spectroscopy (ICP-OES, Optima 8300, PerkinElmer) under standard operating conditions. Samples were digested in concentrated \ce{HNO3} and quantified using external calibration with TruQms multi-element calibration standards. Further analysis of sample compositions were performed using a Zeiss Gemini SEM-EDX with 10 kV accelerating voltage and 200 pA accelerating current at an average 8.9 mm working distance. Further details regarding the crystals and their doping concentrations are given in Supplementary Note 1.

Magnetisation as function of temperature and applied magnetic field was measured using a Quantum Design MPMS-XL SQUID magnetometer in DC mode. Temperature-dependent magnetisation was measured using both zero-field cooling (ZFC) and field-cooling (FC) approaches with an applied magnetic field of 1000 Oe. 

Single crystal neutron diffraction measurements were performed at the Paul Scherrer Institute (Villigen, Switzerland) using the thermal neutron single crystal diffractometer ZEBRA~\cite{zebra_neutron_diffractometer}. Measurements were performed on the \textit{x} = 0.08 sample mounted on a custom built Al holder placed in a Eulerian cradle. The sample environment ranged from 10 to 473 K using both a cryostat (10 to 300 K) and furnace (303 to 473 K). The crystal volume was approximately 2.6 mm$^3$. A Ge$_{311}$ monochromator was used with $\lambda = 1.179$ \AA. Data were collected for specific $hkl$ reflections both allowed and forbidden in the $P6_3/mmc$ parent symmetry with ~9.5 seconds counting time per point and 31 points in each rocking curve.  Data reduction was performed using the pyzebra online suite~\cite{zebra_neutron_diffractometer}. Peaks were fitted with a Gaussian peak shape and the resulting integrated intensities subsequently corrected for the Lorentz factor. Integrated intensities were analysed with Mag2Pol~\cite{Qureshi:in5014} using absorption correction based on sample geometry. The magnetic structure was modelled with multidomain orthorhombic $Cmc'm'$ and monoclinic $C2'/c'$ symmetry with $\chi_r^2$ as refinement quality indicator.

Anomalous Hall measurements were performed with a Quantum Design physical properties measurement system (PPMS-Dynacool). Samples were mounted with I$\parallel$c and B$\parallel$$[01 \bar{1} 0]$, using silver conductive epoxy for reliable electrical contacts in cryogenic environment. The resistivity, Hall resistivity, and Hall conductivity were calculated and subtracted using the van der Pauw method. The temperature dependence of the zero-field Hall component was measured using the following method. The sample was first field-cooled to 5 K under a constant magnetic field of $+7$ T ($-7$ T). At 5 K, the field was reduced to 0 T without changing its sign. Then the Hall resistance was measured by warming the sample to 300 K in zero magnetic field. The longitudinal resistivity component was subtracted out by antisymmetrizing the $+7$ T field-cooled and $-7$ T field-cooled data. Further details regarding the geometry of the measurement are given in Supplementary Note 3.

\section{Results and Discussion}

\begin{table*}
\normalsize
\caption{Atomic percentages (at\%) of the elements in each \CrMnSn\ crystal, determined by ICP-OES and normalised to 25 at\% Sn. The site occupancy denotes the total occupancy of the Mn site, including both Mn and Cr. The normalised Cr content is given by $\mathrm{Cr}/(\mathrm{Mn}+\mathrm{Cr})$, and $x$ follows from it directly. The refined $x = 0.15$ crystal was determined once only, so no reproducibility estimate is available. See Supplementary Note 1 for further details regarding the determination of uncertainties. $<$LOD indicates that the amount was below the limit of detection.}
\label{tab:icp_stoichiometry}

\setlength{\tabcolsep}{4pt}
\begin{tabular}{|c|c|c|c|c|}
\hline
\makecell[l]{Cr \\ (at\%)} &
\makecell[l]{Mn \\ (at\%)} &
\makecell[l]{Cr/Mn site \\ occupancy (\%)} &
\makecell[l]{normalised \\ Cr content (\%)} &
\makecell[l]{refined \\ $x$}\\
\hline\hline
 $<$LOD             & $75.52 \pm 0.06$ & $100.44 \pm 0.24$ & $<$LOD             &  0.00 \\ \hline
 $3.31 \pm 0.12$  & $72.17 \pm 1.08$ & $100.64 \pm 1.29$ & $4.39 \pm 0.21$  & 0.04 \\ \hline
 $3.55 \pm 0.06$  & $71.46 \pm 0.24$ & $100.02 \pm 0.38$ & $4.74 \pm 0.06$  & 0.05 \\ \hline
 $6.16 \pm 0.39$  & $68.50 \pm 0.10$ & $99.54 \pm 0.57$  & $8.25 \pm 0.48$  & 0.08 \\ \hline
 $9.81 \pm 1.54$  & $65.05 \pm 0.24$ & $99.82 \pm 1.74$  & $13.10 \pm 1.83$ & 0.13 \\ \hline
$11.16$           & $63.71$           & $99.83$            & $14.91$           & 0.15 \\ \hline
\end{tabular}
\end{table*}

\noindent High quality single crystals of \CrMnSn\ with varying \textit{x} were obtained, with crystal sizes measuring between 1.8 and 3.5 mm in length and 0.7 to 1.7 mm in diameter, as shown in Figure S1 in Supplementary Note 1. Table \ref{tab:icp_stoichiometry} lists the results from elemental analysis using ICP-OES. The near perfect 100 \% occupancy of the Cr/Mn site in each of the crystals indicates that the Bi-flux method successfully provides stoichiometric 3:1 site occupancies in our \CrMnSn\ crystals, with Cr directly substituting on the Mn site. Note that, in contrast to more commonly employed SEM-EDX, ICP-OES is a bulk-sensitive technique which determines the stoichiometry of full crystal chunks by dissolving them in strong acid, providing accurate bulk stoichiometry. Results obtained by SEM-EDX are listed in Table S2 and shown in Figure S2 in Supplementary Note 1, and are within agreement of our ICP-OES data.  Note that for simplicity, rounded values of \textit{x} are used throughout this work.

\begin{figure}[b]
\centering
\includegraphics[width=0.45\textwidth]{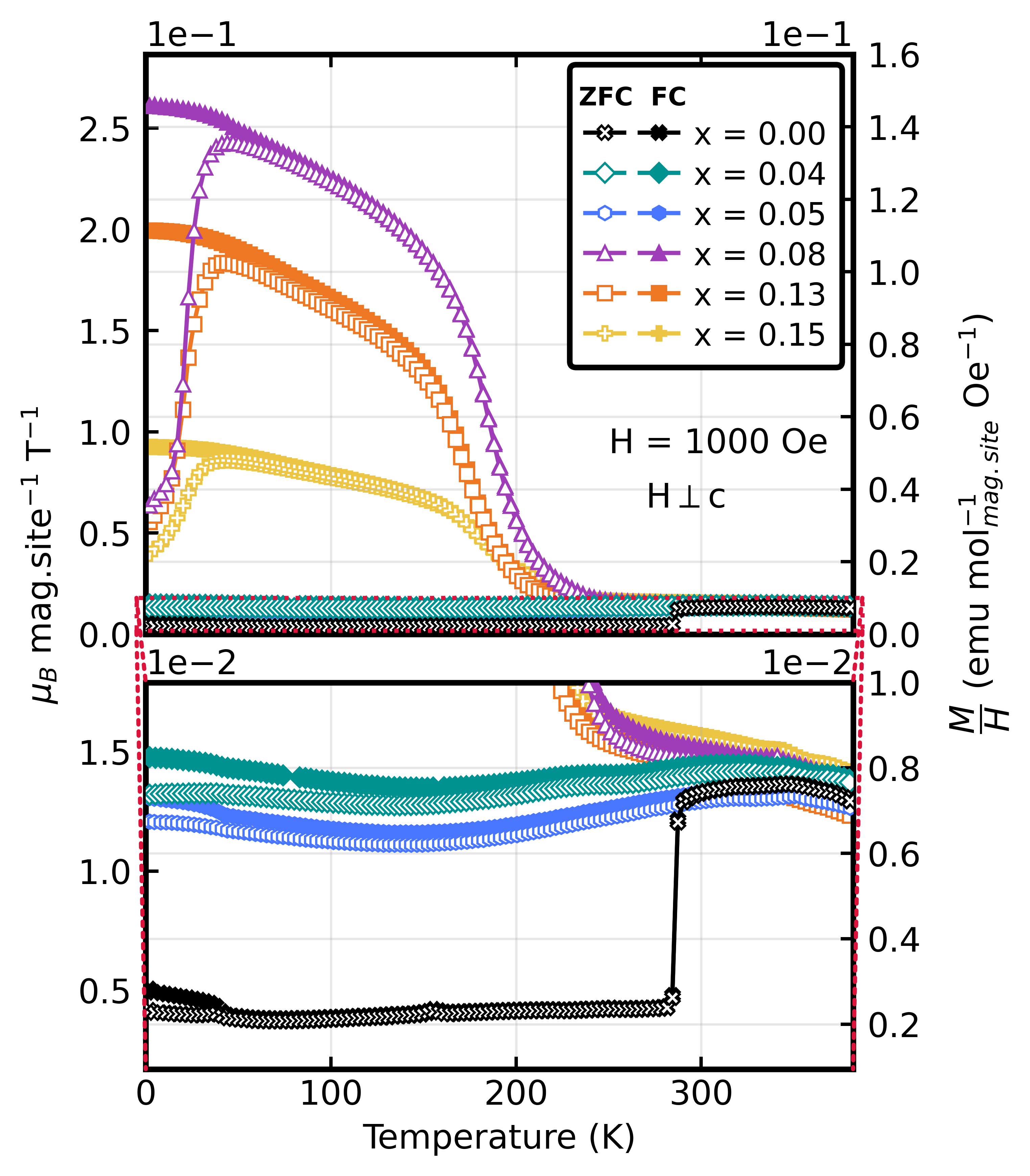}
\caption{Magnetic susceptibility of single crystals of \CrMnSn\ with varying \textit{x}, measured with applied magnetic field of 1000 Oe, using zero-field cooled (open symbols) and field-cooled (full-colour symbols) protocols. Selected measurements with $H \parallel c$ are presented in Figure S4 in Supplementary Note 2. The upper plot shows the full dataset while the lower plot zooms in to capture the magnetic behaviour of samples with lower Cr doping content. }
\label{fig:SQUID_Tsweeps}
\end{figure}

The magnetic susceptibility as a function of temperature is shown in Figure \ref{fig:SQUID_Tsweeps}. Each doping concentration shows distinct magnetic behaviour that can be divided into three subgroups, based on their Cr concentration: undoped ($x = 0$), low-doped $x = 0.04 - 0.05$, and high-doped $x = 0.08 - 0.15$. The stoichiometric undoped sample exhibits well-established magnetic behaviour of the transition from the IT-AFM phase to the helical incommensurate magnetic phase at 290 K \cite{Nagamiya1982, Cable1993,cederholm2026ground}. Moreover, a small anomaly observed near 150 K indicates restructuring in the $ab$-plane \cite{Duan2015, Sung2018} and is unobserved in \textnormal{H}\(\parallel\)\textnormal{c} susceptibility measurements of Figure S4(a) in Supplementary Note 2. At lower temperatures, the onset of a spin-glass phase is seen near 40 K \cite{tomiyoshi1986triangular}, as is apparent from the splitting of the ZFC and FC data, with a slight FM transition for FC only \cite{Feng2006}. These observations are in great agreement with other flux-grown \ce{Mn3Sn} crystals \cite{Sung2018, Park2025}.

The second subgroup corresponds to low-doped \CrMnSn\ with $x = 0.04-0.05$ and shows no sign of a transition to the incommensurate phase. In other words, the IT-AFM phase is stabilised down to the onset of spin glass behaviour near 40 K. This stabilisation behaviour was similarly reported in the parent compound using Mn-rich growth methods such as the Czochralski method \cite{Nakatsuji2015}, yielding \ce{Mn_{3$+$\delta}Sn}. However, based on our elemental analysis (see Table \ref{tab:icp_stoichiometry}), we find that the low-doped \CrMnSn\ crystals have close to perfect 3:1 stoichiometry. This means that our results prove that the IT-AFM phase can also be stabilised in purely stoichiometric \CrMnSn\ crystals, by small amounts of Cr doping on the Mn site, excluding Mn substitution on the Sn site as key attribute to the mechanism of AHE stabilisation~\cite{liu2026giant}. As stated above, while the work of Yano \textit{et al.} also shows this stabilisation effect for a similar Cr doping concentration (\ce{Mn_{2.888}Cr_{0.112}Sn}) \cite{Yano2024}, their crystal is grown via the Bridgeman method and their elemental analysis shows that their crystals contain a small surplus of Mn. 

The third subgroup corresponds to high-doped \CrMnSn\ with $x = 0.08 - 0.15$. For these crystals, the incommensurate phase transition at 290 K is again suppressed. Furthermore, a small FM transition abruptly appears below 250 K.  This change signals symmetry breaking of the robust underlying IT-AFM phase which now includes a stronger FM component due to Cr substitution. Apparent in the magnetisation measurements of Figure S5, found in Supplementary Note 2, is that this symmetry change is field induced and the small remnant magnetisation of the IT-AFM phase is restored at zero field. Therefore the FM transition is observed due to our susceptibility measurements being performed in 1000 Oe field. Most notably, the FM contribution is largest in the $x = 0.08$ crystal. We look to the recent work on Mn rich systems of Liu \textit{et al.} \cite{liu2026giant} to hypothesise that Cr doping at this concentration may similarly have a perturbative effect in inducing a net magnetisation. Below $x = 0.08$, Cr only disrupts the stability of the incommensurate phase, but above it, Cr exchange  near 180 K and with external field can induce symmetry breaking, similar behaviour as reported in Fe doped analogues \cite{Hori1990}. 

\begin{figure*}[htp!]
\centering
\includegraphics[width= 1.0\textwidth]{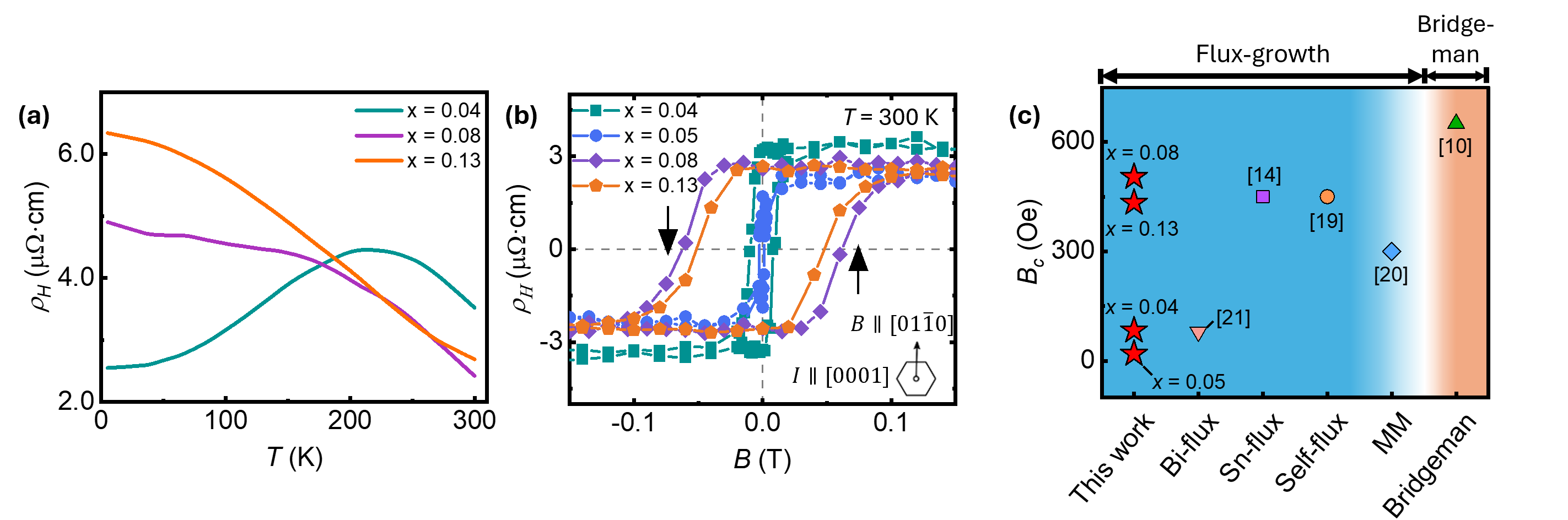}
\caption{(a) Anomalous Hall resistivity as a function of temperature for \CrMnSn\ with various $x$. (b) Hall resistivity of \CrMnSn\ with various $x$, measured at room temperature ($T=$ 300 K) with $B\parallel[01\bar 1 0]$ and $I\parallel[0001]$. Field sweep directions are indicated by arrows. (c) Coercive field ($B_c$) values of \CrMnSn\ synthesized by various methods, as reported in literature and in this work. MM in the horizontal axis stands for Molten metal flux. The shaded regions indicatee different synthesis methods: left: flux methods, right: Bridgman method.}
\label{fig:AHE_RT}
\end{figure*}

The anomalous Hall signature of the second and third subgroup of samples, shown in Figure \ref{fig:AHE_RT}(a) and (b), displays distinct transport behaviour. Figure \ref{fig:AHE_RT}(a) shows the AHE as a function of temperature in the $x = $ 0.04, 0.08, and 0.13 crystals. The second subgroup, sample $x=0.04$, that hosts a stabilized IT-AFM phase down to lowest temperature, shows strong non-monotonic AHE characteristics down to 5 K. This comes from intrinsic Berry curvature \cite{Yano2024,gas2025magnetic}, and the non-monotonic temperature dependence matches with the behaviour reported in a similarly doped sample by Yano et al.  \cite{Yano2024}. 

In both third subgroup crystals ($x=0.08$ and $x=0.13$), the AHE signal increases monotonically with decreasing temperature. This is consistent with their temperature-dependent magnetic susceptibility behaviour \cite{Yano2024}. Importantly, increasing the doping from $x =$ 0.08 to $x =$ 0.13 leads to a reduction of susceptibility, but an increase of AHE. This discrepancy could result from either the fact that these two measurements were performed at two different fields ($B=0$ T for AHE and $B=0.1$ T for susceptibility), or a slight misorientation of the $x=0.13$ sample because of its irregular shape.

The magnetic field-dependent anomalous Hall measurements, shown in Figure \ref{fig:AHE_RT}(b), were performed on \CrMnSn\ with different Cr doping concentrations at $T = $ 300 K.  $B_c$ and anomalous resistivity ($\rho_H$) from all the samples are listed in Table S5 in Supplementary Note 3. Samples from both subgroup 2 and 3 show a significant spontaneous Hall resistivity between 2.3 to 3.3 $\mu\Omega\cdot\text{cm}$. This magnitude matches with previous reports~\cite{nakatsuji2015large, Yano2024}. 

\begin{figure*}[t!]
    \centering
    \includegraphics[width=1.0\textwidth]{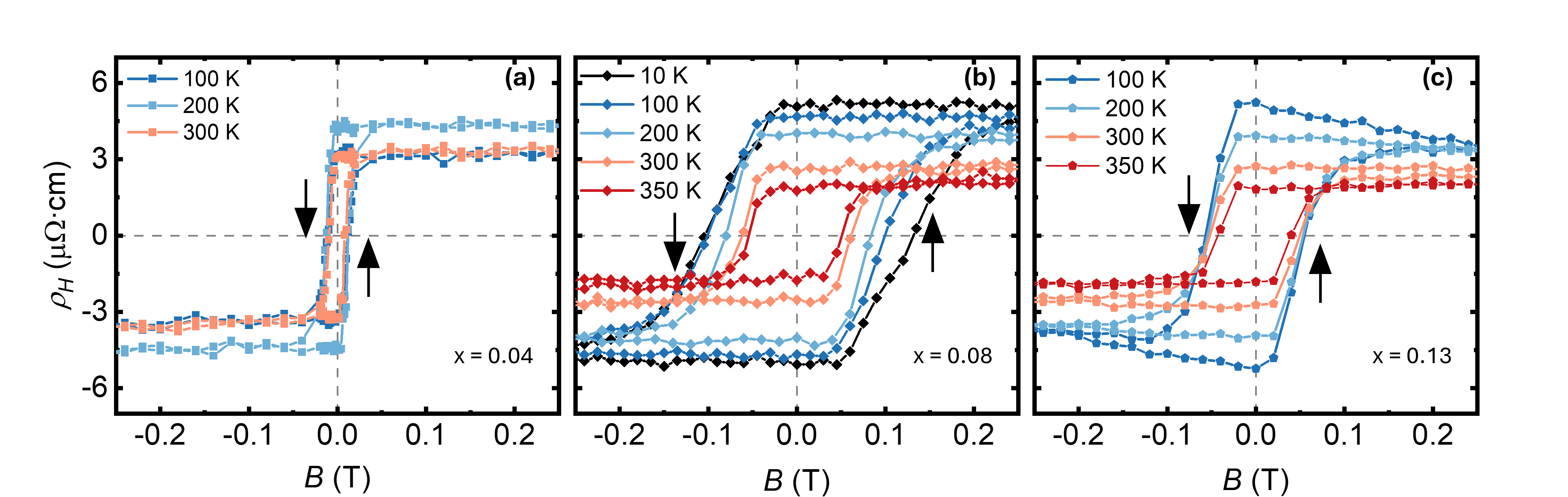}
    \caption{Temperature-dependent Hall resistivity of \CrMnSn\ with \textbf{(a)} $x = 0.04$, \textbf{(b)} $x = 0.08$, \textbf{(c)} $x = 0.13$, in $B\parallel[01\bar 1 0]$ with $I\parallel[0001]$. Field sweep directions are indicated by arrows.}
    \label{fig:PPMS_Graphs}
\end{figure*}

 A summary of $B_c$ values of \ce{(MnX)_{3}Sn} crystals synthesized by us and reported by others is shown in Figure \ref{fig:AHE_RT}(c). In most cases, $B_c$ values range from 300 - 700 Oe other than $B_c=80$ Oe reported by Nakatsuji et al.~\cite{nakatsuji2015large} in self-doped crystals (containing excess Mn). Our high doped crystals ($x=$ 0.08 and 0.13) have $B_c$ values in the same range. In contrast, we find that small amounts of Cr doping, $x=$ 0.04 and 0.05, in stoichiometric \ce{Mn3Sn} results in $B_c$ values of 80 Oe and 20 Oe respectively, making the latter at least two-thirds smaller than those of native \ce{Mn3Sn}, and the lowest coercivity reported in this compound so far to our knowledge. These findings make low Cr doped \ce{Mn3Sn} exceptionally suitable for low-energy spintronic applications, as the reduced critical field enables more efficient switching, while also preserving the IT-AFM phase across the whole temperature range, before the onset of the spin glass transition (see Figure \ref{fig:SQUID_Tsweeps}). Moreover, we note that recent first-principles calculations by Zhang \textit{et al.} on Cr-doped \ce{Mn3Sn} pinpoint $x=0.0417$ as a unique composition with increased anomalous Hall conductivity, that they argue come from Berry curvature hotspots being in closer alignment with the Fermi level as compared to pristine \ce{Mn3Sn} \cite{Zhang2026_Cr}. We argue that this could also be the underlying cause for the behaviour we observe in the same doping regime.

 A more detailed field dependence of the AHE at various temperatures for crystals $x=$ 0.04, 0.08, and 0.13 is shown in Figure \ref{fig:PPMS_Graphs} which highlights further distinctions between doping behaviours. Figure \ref{fig:PPMS_Graphs}(a) shows the non-monotonic temperature dependence of $\rho_H$ in $x=0.04$ as noted previously in Figure \ref{fig:AHE_RT}(a). The $B_c$ value, however, remains the same ($80$ Oe) for all three temperatures. Notably, the anomalous Hall behaviour of this sample is very similar to previously reported slightly self-doped Mn$_3$Sn sample \cite{nakatsuji2015large} emphasising that the anomalous Hall effect in this sample has an intrinsic origin tied to Berry curvature. 

Both $x = 0.08$ (Fig. \ref{fig:PPMS_Graphs}(b)) and $x=0.13$ (Figure \ref{fig:PPMS_Graphs}(c)) show slanted hysteresis loops indicating either severe domain pinning or alignment of the magnetic field along a hard axis.  Additionally, both of these crystals show monotonic temperature dependence of Hall resistivity as observed previously in Figure \ref{fig:AHE_RT}(a). However, there is a notable difference. Whereas the $B_c$ increases slightly with decreasing temperature in $x=0.13$, $x=0.08$ shows a sharp increase from roughly 500 Oe at 350 K to twice that value at 100 K. It remains unclear what drives the temperature dependence of $B_c$. However, evidence of some similar behaviour has been observed in the Co doped analogue by Zhang \textit{et al.} \cite{Zhang2026_Co}. Additionally, this sample also demonstrates an exchange bias at 10 K indicating a coexistence of FM and AFM domains. These results indicate that the $x = 0.08$ crystal remains unique and we argue that it may lie close to a phase boundary of some kind. Thus studies of additional compounds close to this value and that of $x = 0.05$, which has the smallest $B_c$, would be necessary to pinpoint this behaviour. 

It is important to note that $x=0.13$ shows a high field slope which we attribute to an ordinary Hall effect component. In general, the ordinary Hall component is absent in all other crystals we measured owing to the low-field range chosen for these measurements. As per Table S6 and methods outlined in Supplementary Note 4, we estimate a Hall carrier density of $10^{20}$ cm$^{-3}$ indicating a lower density than a typical \ce{Mn3Sn} system \cite{Nakatsuji2015,Li2017}. This observation of lower density for $x=0.13$ can be attributed to Fermi-level tuning by chemical doping \cite{Zhang2026_Cr}.

While the magnetic phases and anomalous transport in this doping scheme are inextricably linked, we emphasise that the sign of $B_c$ observed in anomalous transport is decoupled from what is seen in magnetisation measurements (see Figure S5 in Supplementary Note 2). The trend we observe in remnant magnetisation in the $ab$ plane for the $x=0.08$ and $x=0.13$ crystals show that zero field magnetisation is similar at both 300 K and 125 K, indicating a restorative tendency toward the IT-AFM. This experimental observation is supported by the calculations from Zhang \textit{et al.}, who state that their calculations converged back to states very close to the coplanar limit, i.e. with all spins in the $ab$-plane \cite{Zhang2026_Cr}. However, $B_c$ observed in our AHE measurements is orders of magnitude larger and exhibits the aforementioned temperature dependence trends. We therefore confirm that AHE in this material is decoupled from the sign of the magnetisation up to a certain threshold value and is not immediately driven by it. This observation confirms the origin of the AHE as an independent band structure property and not resulting directly from the magnetisation within this material \cite{chen2014anomalous}, in agreement with conclusions drawn by Zhao \textit{et al.} for thin films of \ce{Mn3Sn} \cite{Zhao2021}. 

\begin{figure*}[t!]
    \centering
    \includegraphics[width=\textwidth]{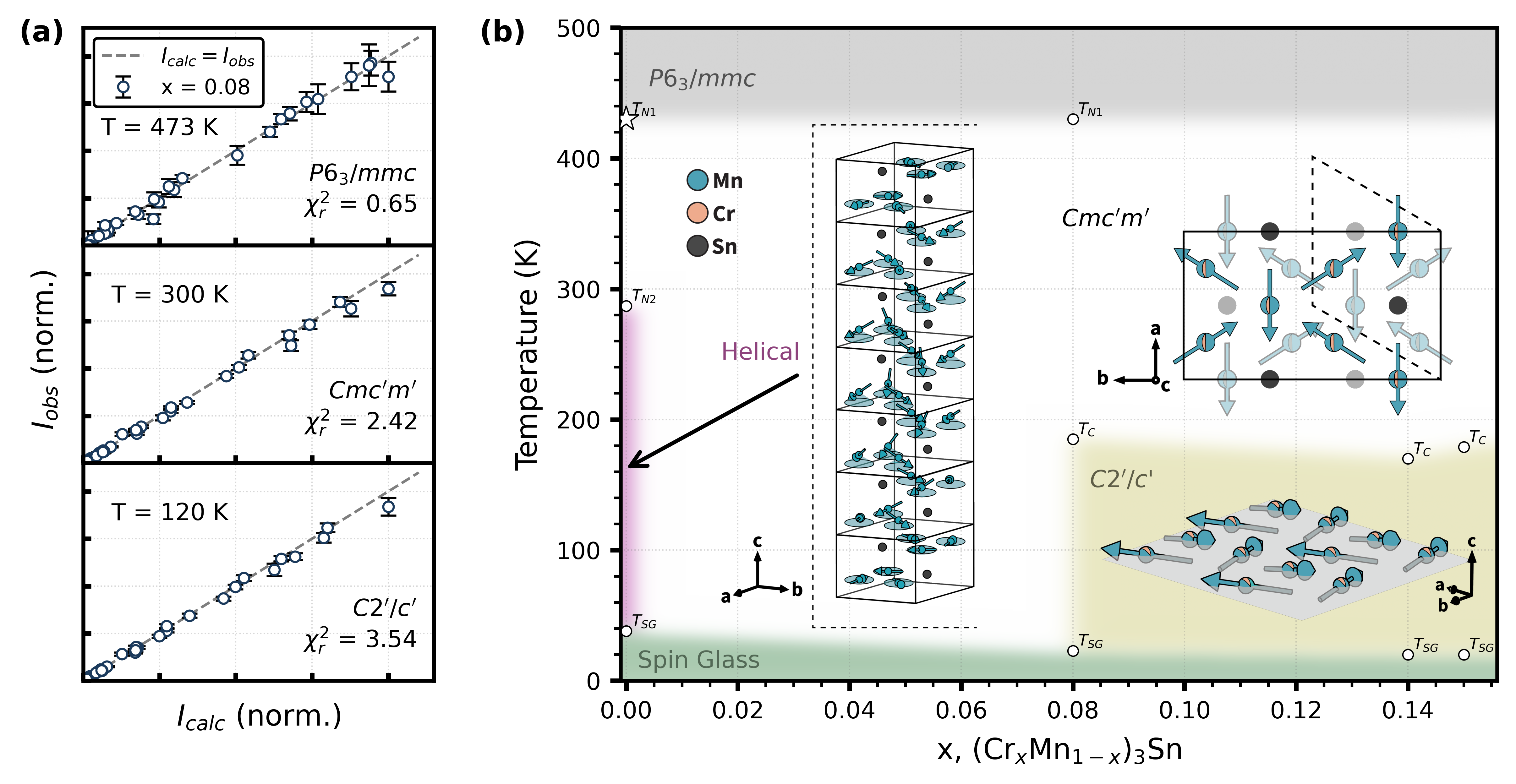}
    \caption{\textbf{(a)} Fits of observed intensities ($I_{\rm obs}$) and calculated intensities ($I_{\rm calc}$) for the neutron diffraction experiment on \ce{(Cr_{0.08}Mn_{0.92})_{3}Sn}, measured at 473 K, 300 K and 120 K using both nuclear and magnetic reflections. We attribute the $<1\chi_r^2$ for the paramagnetic phase to be due to conservative errors of the refinement. \textbf{(b)} Schematic magnetic phase diagram for \CrMnSn\ as a function of temperature and $x$. Shaded regions represent the different magnetic phases (unshaded being $Cmc'm'$), with structures and spacegroups indicated. Open circles represent transition temperatures observed in magnetic susceptibility and neutron diffraction data of this work. The open star, $T_{N1}$ at $x = 0$, is taken from \cite{kouvel1965exchange}, all other transition temperatures are determined in this work. Note that the out-of-plane spin canting in $C2'/c'$ is exaggerated for visibility purposes, and is only present in case of an applied magnetic field.}
    \label{fig:Phase_diagram}
\end{figure*}

In order to gain more insight into the ferromagnetism present in the high-doped samples, we performed single crystal neutron diffraction on the $x = 0.08$ crystal, i.e. the one with the largest FM contribution. As shown by the refinement in Figure \ref{fig:Phase_diagram}(a), \ce{(Cr_{0.08}Mn_{0.92})_3Sn} adopts the $P6_3/mmc$ spacegroup in the paramagnetic state, measured at 473 K. This is the same as native \ce{Mn3Sn} \cite{kren1975study,tomiyoshi1982polarized}. In undoped \ce{Mn3Sn}, the system develops non-collinear AFM order with propagation vector \textbf{k} = 0. Herein, the neighbouring Mn spins in the kagom\'e planes adopt a 120° arrangement, accompanied by a small FM moment of 0.002-0.003 $\mu_\text{B}$ per Mn \cite{nakatsuji2015large, Cable1993, Zimmer1972, tomiyoshi1982polarized} confined within the basal plane. This drives a reduction from hexagonal $P6_3/mmc$ to orthorhombic $Cmc'm'$ symmetry. Here we find that the Cr-doped crystal adopts the same $Cmc'm'$ magnetic space group of the IT-AFM phase at 300 K, with only a slight \textit{ab}-plane FM component (0.00105(5)$\mu_B$ per Mn) observable in field-dependent magnetisation of Figure S5. Choosing 120 K to measure neutron diffraction where the Cr induced FM contribution is enhanced, we find the magnetic symmetry is very well described by the magnetic space group $C2'/c'$ with goodness of fit parameter of $\chi_r^2=3.54$ shown in Table \ref{tab:params}. Although $Cmc'm'$ also describes the data well at this temperature, it does not accommodate the out-of-plane magnetic component that magnetisation measurements show can be induced in applied field (see Figure \ref{fig:SQUID_Tsweeps}). Since neutron diffraction measurements performed on ZEBRA were conducted in zero field, these out-of-plane components are too small to be determined from the neutron data, resulting in correspondingly large uncertainties. Nevertheless, we conclude using values obtained by field-dependent magnetisation measurements shown in Figure S5. Specifically, the magnitude of \textnormal{H}\(\parallel\)\textnormal{c}  at 125 K and 5 K are in the order of 0.00106(7) and 0.01206(11) $\mu_B$ per Mn respectfully. The additional degrees of freedom afforded by the $C2'/c'$ description become increasingly stable at zero field with decreasing temperature. However, the increase in $\chi_r^2$ at 10 K in our refinement model indicates that $C2'/c'$ no longer provides an adequate description of the magnetic structure as the system enters the spin glass phase.   
\begin{table}[ht]
\centering
\begin{threeparttable}
\caption{Structural and magnetic parameters of Cr-doped Mn$_3$Sn from single-crystal
neutron diffraction (ZEBRA, $\lambda = 1.178$~\AA) using Mag2Pol software \cite{Qureshi:in5014}. Mn/Cr occupy the $6h$ site
$(x, 2x, \tfrac14)$ and Sn the $2c$ site $(\tfrac13, \tfrac23, \tfrac14)$, given in the
parent hexagonal cell. Three populated domains fixed at equal fractions; $\tfrac13,\tfrac13,\tfrac13$, the remaining three fixed at zero.}
\label{tab:params}
\small
\begin{tabular}{l|ccc}
\toprule
$T$ (K)                                  & 300              & 120               & 10               \\
MSG                           & $Cmc'm'$           & $C2'/c'$            & $C2'/c'$           \\
$\chi_r^2$                       & 2.42              & 3.54               & 8.05              \\
\midrule
$x$ (Mn/Cr)                              & 0.8389(5)        & 0.8386(6)         & 0.8383(8)        \\
$B_{\mathrm{Mn/Cr}}$ (\AA$^2$)  & 1.0(3)           & 0.6(3)            & 0.6(5)           \\
$B_{\mathrm{Sn}}$ (\AA$^2$)              & 0.9(2)           & 0.4(3)            & 0.4(5)           \\
\midrule
$\mu_{ab}$ ($\mu_\mathrm{B}$)   & 2.2(3)           & 2.6(4)            & 2.7(7)           \\
$\mu_{c}$ ($\mu_\mathrm{B}$)             & 0.0*       & undetermined & undetermined           \\
\bottomrule
\end{tabular}
\begin{tablenotes}\footnotesize
\item[*] * Value confined by symmetry.
\end{tablenotes}
\end{threeparttable}
\end{table}
The $C2'/c'$ magnetic space group can be described by a superposition of the irreducible representations mGM2+ and mGM6+. Using this symmetry allows for more spin degrees of freedom than the $mm'm'$ magnetic point group of the IT-AFM phase where spin rotations are confined to an axis, while $2'/m'$ confines spins to a mirror plane, essential to facilitating the net magnetisation observed in magnetic susceptibility induced with external field. These magnetic structures are summarised schematically in the magnetic phase diagram of Figure \ref{fig:Phase_diagram}(b). This, providing a framework to better understand the interplay of topology and magnetism in Cr-doped \ce{Mn3Sn} systems.

\section{Conclusion}
\noindent To summarize, we have shown that systematic Cr doping in (Cr$_x$Mn$_{1-x}$)$_3$Sn with pure 3:1 (Mn+Cr):Sn stoichiometry provides a powerful route to tuning both the magnetic properties and the anomalous Hall response of this prototypical Weyl semimetal. In particular, low doping concentrations of $x = 0.04 - 0.05$ suppress the competing incommensurate phase of native \ce{Mn3Sn}. This behaviour, characteristic of excess Mn synthesis routes, can now also be attributed directly to the dopant which in addition  significantly reduces the $B_c$ required for anomalous Hall switching. These results establish Cr doping as an effective control parameter for reducing dissipation in future AFM devices.

At $x = 0.08$, a distinct field induced magnetic transition emerges near 180 K, accompanied by a symmetry reduction from the IT-AFM $Cmc'm'$ phase to $C2'/c'$ that allows for spin canting out of the $ab$-plane. The AHE with this composition exhibits pronounced temperature dependence of $B_c$ for anomalous Hall switching, distinguishing it from both higher and lower doping regimes. This highlights a unique coupling between $x = 0.08$ doping and transport properties that we argue could stem from its position on a phase boundary.

These results provide new insight into the interplay between AHE and magnetic behaviour in \ce{Mn3Sn} based systems, while showing that chemical doping explicitly can provide a practical handle for engineering their functional properties. More broadly, this work highlights the potential of Cr-doped \ce{Mn3Sn} for spintronic applications, where reduced switching fields and tunable Hall responses are highly desirable.

\section*{acknowledgments}
\noindent This work is based on experiments performed at the Swiss spallation neutron source SINQ, Paul Scherrer Institute, Villigen, Switzerland (proposal numbers 20240942 and 20241604). P.M. and M.E.K. acknowledge the research program \textit{Materials for the Quantum Age} (QuMat) for financial support. This program (registration number 024.005.006) is part of the Gravitation program financed by the Dutch Ministry of Education, Culture and Science (OCW). This work was supported by the Danish Committee for Research Infrastructure (NUFI) through Q-MAT and DanScatt. H.J. was funded by the Carlsberg Foundation CF200497. We thank Keld Theodor and the rest of the technicians at KU for making the sample holder for the neutron scattering experiment.

\bibliographystyle{ieeetr}
\bibliography{bibliography}

\end{document}


\title{Supplementary Information: \\ Magnetic phase evolution and Hall effect modulation in Cr-doped stoichiometric \ce{Mn3Sn}}

\author{Padraig Maderson}
\affiliation{Debye Institute for Nanomaterials Science, Utrecht University, 3508 TA Utrecht, the Netherlands}
\author{Zhiyuan Cheng}
\affiliation{Leiden Institute of Physics, Leiden University, Niels Bohrweg 2, 2333 CA Leiden, the Netherlands}
\author{Iris Serra} 
\affiliation{Leiden Institute of Physics, Leiden University, Niels Bohrweg 2, 2333 CA Leiden, the Netherlands}
\author{Romain Sibille}
\affiliation{PSI Center for Neutron and Muon Sciences, CH-2532 Villigen-PSI, Switzerland}
\author{Jiacheng Zhang} 
\affiliation{State Key Laboratory of Quantum Functional Materials, School of Physical Science and Technology, ShanghaiTech University, Shanghai 201210, China}
\author{Yangfeng Guo} 
\affiliation{State Key Laboratory of Quantum Functional Materials, School of Physical Science and Technology, ShanghaiTech University, Shanghai 201210, China}
\affiliation{ShanghaiTech Laboratory for Topological Physics, ShanghaiTech University, Shanghai 201210, China}
\author{Navid Qureshi}
\affiliation{Institut Laue-Langevin, 71, avenue de Martyrs, 38021 Grenoble, France}
\author{Semonti Bhattacharyya}
\affiliation{Leiden Institute of Physics, Leiden University, Niels Bohrweg 2, 2333 CA Leiden, the Netherlands}
\author{Henrik Jacobsen}
\email{henrik.jacobsen@ess.eu}
\affiliation{Nanoscience Center, Niels Bohr Institute, University of Copenhagen, 2100 Copenhagen, Denmark}
\affiliation{Data Management and Scientific Computing, Asmussens Allé 305, 2800 Kongens Lyngby, Denmark}
\author{Machteld E. Kamminga}
\email{m.e.kamminga@uu.nl}
\affiliation{Debye Institute for Nanomaterials Science, Utrecht University, 3508 TA Utrecht, the Netherlands}

\maketitle

\raggedbottom

\section*{Supplementary Note 1: sample information}

\begin{figure}[H]
    \centering     \includegraphics[width=0.9\textwidth]{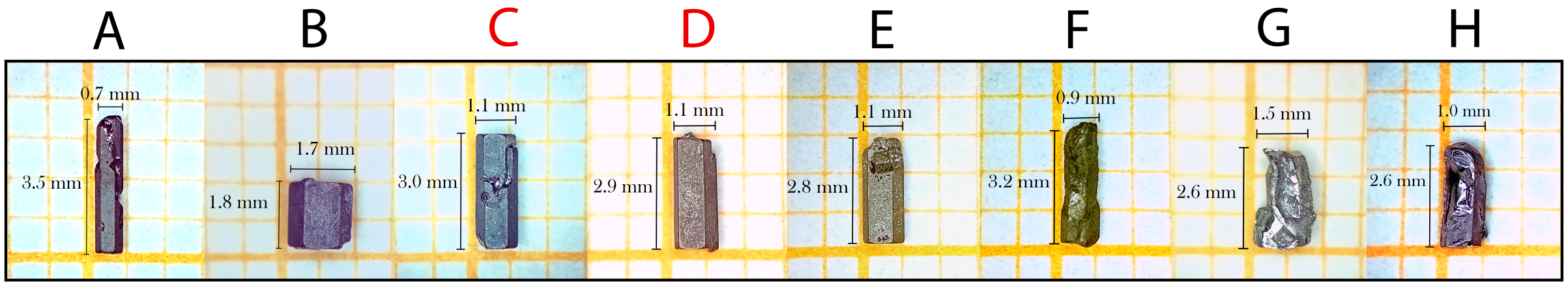}
    \caption{Photograph of selected crystals. Crystals C and D (red) have similar Cr concentrations to E and were therefore left out of the main text.}
    \label{fig:crystals}
\end{figure}

\noindent A total of eight crystals were synthesised, with different target Cr concentrations (see Figure S1 and Table S1) via the method outlined in the main text. Table S1 shows the atomic percentages of each crystal as determined by ICP-OES. As shown in Table S1, the actual Cr concentrations all are larger than the nominal target concentration, indicating that Cr might preferentially migrate towards the crystal. In addition, we note that crystals C and D have very similar Cr concentrations as E and are therefore excluded from further studies.

The error associated with ICP-OES is largest due to inconsistencies in sample digestion between runs. Some values in Table S2 have been marked with a dagger and removed from the calculation of averages, as poor digestion in \ce{HNO3} was the likely cause. 

SEM-EDX measurements were also performed, as shown in Table S2. We note that SEM-EDX is a surface-sensitive technique, while ICP-OES is a bulk technique. The SEM-EDX measurements involved point spectrums on specific regions of the whole crystal for which crystal E, with $x=0.05$, had a Bi inclusion observed in Figure S2 and Table S4 due to the Bi flux synthesis. The following measurements were then made on cross-sections of the crystals cut via steel razor and these did not show any evidence that Bi alloys in any of the crystals, yielding proper agreement with ICP-OES, as shown in Table S3.

\begin{table}[htbp]
\small
\caption{Atomic percentages (at\%) of the elements in each \CrMnSn\ crystal, determined by ICP-OES and normalised to 25 at\% Sn. Values are means over $N$ independent determinations, excluding those flagged in Table~\ref{tab:all_determinations}. The site occupancy denotes the total occupancy of the Mn site, including both Mn and Cr. The normalised Cr content is given by $\mathrm{Cr}/(\mathrm{Mn}+\mathrm{Cr})$, and $x$ follows from it directly. Uncertainties are $2\sigma$ standard errors of the mean across determinations; these exceed the analytical precision of any single measurement by two to three orders of magnitude and therefore represent the reproducibility of the full digestion-and-calibration procedure. Crystal H was determined once only, so no reproducibility estimate is available. $<$LOD indicates that the amount was below the limit of detection.}
\label{tab:icp_stoichiometry}

\setlength{\tabcolsep}{4pt}
\begin{tabular}{|c|c|c|c|c|c|c|}
\hline
\makecell[l]{Crystal} &
\makecell[l]{nominal \\ Cr (\%)} &
\makecell[l]{$N$} &
\makecell[l]{Cr \\ (at\%)} &
\makecell[l]{Mn \\ (at\%)} &
\makecell[l]{Cr/Mn site \\ occupancy (\%)} &
\makecell[l]{normalised \\ Cr content (\%)} \\
\hline\hline
A & 0.0  & 2 & $<$LOD             & $75.515 \pm 0.064$ & $100.444 \pm 0.243$ & $<$LOD             \\ \hline
B & 1.5  & 3 & $3.309 \pm 0.121$  & $72.171 \pm 1.075$ & $100.639 \pm 1.285$ & $4.385 \pm 0.211$  \\ \hline
C & 2.0  & 2 & $3.556 \pm 0.054$  & $71.477 \pm 0.247$ & $100.045 \pm 0.400$ & $4.739 \pm 0.053$  \\ \hline
D & 3.0  & 2 & $3.532 \pm 0.301$  & $71.368 \pm 0.185$ & $99.867 \pm 0.648$  & $4.716 \pm 0.371$  \\ \hline
E & 4.5  & 3 & $3.553 \pm 0.056$  & $71.458 \pm 0.237$ & $100.015 \pm 0.383$ & $4.736 \pm 0.058$  \\ \hline
F & 5.0  & 3 & $6.156 \pm 0.391$  & $68.495 \pm 0.095$ & $99.536 \pm 0.573$  & $8.245 \pm 0.478$  \\ \hline
G & 8.0  & 2 & $9.812 \pm 1.544$  & $65.053 \pm 0.239$ & $99.820 \pm 1.740$  & $13.099 \pm 1.833$ \\ \hline
H & 11.0 & 1 & $11.161$           & $63.712$           & $99.831$            & $14.906$           \\ \hline
\end{tabular}
\end{table}

\begin{table}[H]
\centering
\small
\caption{Atomic percentages (at\%) of the elements in each \CrMnSn\ crystal, from three independent ICP-OES determinations using the site-based calibration and from SEM-EDX, normalised to 25 at\% Sn. The site occupancy denotes the total occupancy of the Mn site, including both Mn and Cr. The normalised Cr content is given by $\mathrm{Cr}/(\mathrm{Mn}+\mathrm{Cr})$. ICP-OES uncertainties are $2\sigma$ analytical precision and do not include systematic contributions from digestion or calibration; SEM-EDX uncertainties are $2\sigma$ standard errors over $N$ spectra. Entries marked $^\dagger$ were excluded from the combined values of Table~\ref{tab:combined_cr_content} on the grounds of being less than the limit of detection (LOD), anomalous $2\sigma$, or site occupancy deviating from 100\% by more than 5 percentage points.}
\label{tab:all_determinations}
\setlength{\tabcolsep}{3pt}
\begin{tabular}{|c|c|c|c|c|c|c|}
\hline
\makecell[l]{Crystal} &
\makecell[l]{nominal \\ Cr (\%)} &
\makecell[l]{$N$} &
\makecell[l]{Cr \\ (at\%)} &
\makecell[l]{Mn \\ (at\%)} &
\makecell[l]{Cr/Mn site \\ occupancy (\%)} &
\makecell[l]{normalised \\ Cr content (\%)} \\
\hline\hline
\multicolumn{7}{|l|}{\textbf{ICP-OES Run 1}} \\ \hline
A & 0.0  & --- & $<$LOD            & $75.483 \pm 0.004$ & $100.565 \pm 0.005$ & $<$LOD$^\dagger$   \\ \hline
B & 1.5  & --- & $3.201 \pm 0.001$ & $73.240 \pm 0.006$ & $101.922 \pm 0.007$ & $4.188 \pm 0.002$  \\ \hline
E & 4.5  & --- & $3.601 \pm 0.001$ & $71.575 \pm 0.005$ & $100.235 \pm 0.005$ & $4.790 \pm 0.001$  \\ \hline
F & 5.0  & --- & $6.501 \pm 0.001$ & $68.478 \pm 0.006$ & $99.973 \pm 0.006$  & $8.671 \pm 0.002$  \\ \hline
G & 8.0  & --- & $9.040 \pm 0.002$ & $65.172 \pm 0.005$ & $98.950 \pm 0.005$  & $12.182 \pm 0.003$ \\ \hline
\multicolumn{7}{|l|}{\textbf{ICP-OES Run 2}} \\ \hline
A & 0.0  & --- & $<$LOD             & $75.547 \pm 0.004$ & $100.322 \pm 0.006$ & $<$LOD$^\dagger$   \\ \hline
B & 1.5  & --- & $3.410 \pm 0.001$  & $71.537 \pm 0.005$ & $99.929 \pm 0.006$  & $4.549 \pm 0.001$  \\ \hline
C & 2.0  & --- & $3.583 \pm 0.001$  & $71.601 \pm 0.003$ & $100.245 \pm 0.003$ & $4.766 \pm 0.001$  \\ \hline
D & 3.0  & --- & $3.683 \pm 0.001$  & $71.460 \pm 0.006$ & $100.191 \pm 0.007$ & $4.902 \pm 0.002$  \\ \hline
E & 4.5  & --- & $3.553 \pm 0.001$  & $71.579 \pm 0.007$ & $100.176 \pm 0.008$ & $4.729 \pm 0.002$  \\ \hline
F & 5.0  & --- & $6.142 \pm 0.002$  & $68.585 \pm 0.006$ & $99.637 \pm 0.006$  & $8.219 \pm 0.003$  \\ \hline
G & 8.0  & --- & $10.584 \pm 0.002$ & $64.933 \pm 0.004$ & $100.690 \pm 0.003$ & $14.015 \pm 0.003$ \\ \hline
H & 11.0 & --- & $11.161 \pm 0.002$ & $63.712 \pm 0.004$ & $99.831 \pm 0.003$  & $14.906 \pm 0.003$ \\ \hline
\hline
\multicolumn{7}{|l|}{\textbf{ICP-OES Run 3}} \\ \hline
A & 0.0  & --- & $<$LOD            & $77.633 \pm 0.004$ & $103.328 \pm 0.006$ & $<$LOD$^\dagger$          \\ \hline
B & 1.5  & --- & $3.315 \pm 0.002$ & $71.736 \pm 0.006$ & $100.067 \pm 0.007$ & $4.417 \pm 0.002$         \\ \hline
C & 2.0  & --- & $3.529 \pm 0.002$ & $71.354 \pm 0.007$ & $99.845 \pm 0.009$  & $4.713 \pm 0.002$         \\ \hline
D & 3.0  & --- & $3.382 \pm 0.002$ & $71.275 \pm 0.009$ & $99.543 \pm 0.010$  & $4.531 \pm 0.003$         \\ \hline
E & 4.5  & --- & $3.504 \pm 0.002$ & $71.221 \pm 0.012$ & $99.633 \pm 0.015$  & $4.690 \pm 0.003$         \\ \hline
F & 5.0  & --- & $5.825 \pm 0.004$ & $68.423 \pm 0.014$ & $98.997 \pm 0.015$  & $7.845 \pm 0.007$         \\ \hline
G & 8.0  & --- & $7.011 \pm 0.022$ & $77.428 \pm 0.048$ & $112.585 \pm 0.042$ & $8.303 \pm 0.028^\dagger$ \\ \hline
H & 11.0 & --- & $7.430 \pm 0.022$ & $80.508 \pm 0.046$ & $117.251 \pm 0.037$ & $8.449 \pm 0.027^\dagger$ \\ \hline
\hline
\multicolumn{7}{|l|}{\textbf{SEM-EDX}} \\ \hline
A & 0.0  & 15 & $<$LOD          & $74.013 \pm 0.258$ & 98.684 & $<$LOD           \\ \hline
B & 1.5  & 46 & $2.051 \pm 0.207$ & $71.420 \pm 0.298$ & 97.960 & $2.792 \pm 0.274$ \\ \hline
E & 4.5  & 27 & $3.539 \pm 0.211$ & $69.511 \pm 0.417$ & 97.400 & $4.845 \pm 0.276$ \\ \hline
F & 5.0  & 48 & $5.950 \pm 0.289$ & $67.924 \pm 0.227$ & 98.500 & $8.054 \pm 0.360$ \\ \hline
G & 8.0  & 41 & $8.949 \pm 0.279$ & $62.759 \pm 0.464$ & 95.647 & $12.480 \pm 0.350$ \\ \hline
\end{tabular}
\end{table}

\begin{table}[htbp]
\normalsize
\caption{Normalised Cr content, $\mathrm{Cr}/(\mathrm{Mn}+\mathrm{Cr})$, and refined stoichiometry parameter $x$ for each \CrMnSn\ crystal. Values are the mean of the $N$ independent determinations that passed the quality criteria of Table~\ref{tab:all_determinations}; all ICP-OES determinations use the site-based calibration. Uncertainties are $2\sigma$ across those determinations. The refined $x$ is taken from the combined ICP-OES $+$ SEM-EDX mean.}
\label{tab:combined_cr_content}
\setlength{\tabcolsep}{4pt}
\begin{tabular}{|c|c|c|c|c|c|c|}
\hline
\makecell[l]{Crystal} &
\makecell[l]{nominal \\ Cr (\%)} &
\multicolumn{2}{c|}{\makecell[l]{ICP-OES only}} &
\multicolumn{2}{c|}{\makecell[l]{ICP-OES $+$ SEM-EDX}} &
\makecell[l]{refined \\ $x$} \\
\cline{3-6}
 & & \makecell[l]{norm.\ Cr (\%)} & \makecell[l]{$n$}
   & \makecell[l]{norm.\ Cr (\%)} & \makecell[l]{$n$} & \\
\hline\hline
A & 0.0  & ---                & 0 & $<$LOD             & 1 & $<0.001$ \\ \hline
B & 1.5  & $4.385 \pm 0.365$  & 3 & $3.986 \pm 1.620$  & 4 & 0.040 \\ \hline
C & 2.0  & $4.739 \pm 0.075$  & 2 & $4.739 \pm 0.075$  & 2 & 0.047 \\ \hline
D & 3.0  & $4.716 \pm 0.525$  & 2 & $4.716 \pm 0.525$  & 2 & 0.047 \\ \hline
E & 4.5  & $4.736 \pm 0.101$  & 3 & $4.764 \pm 0.136$  & 4 & 0.048 \\ \hline
F & 5.0  & $8.245 \pm 0.827$  & 3 & $8.197 \pm 0.702$  & 4 & 0.082 \\ \hline
G & 8.0  & $13.099 \pm 2.592$ & 2 & $12.892 \pm 1.967$ & 3 & 0.129 \\ \hline
H & 11.0 & 14.906             & 1 & 14.906             & 1 & 0.149 \\ \hline
\end{tabular}
\end{table}

\begin{figure}
    \centering
    \includegraphics[width=0.98\linewidth]{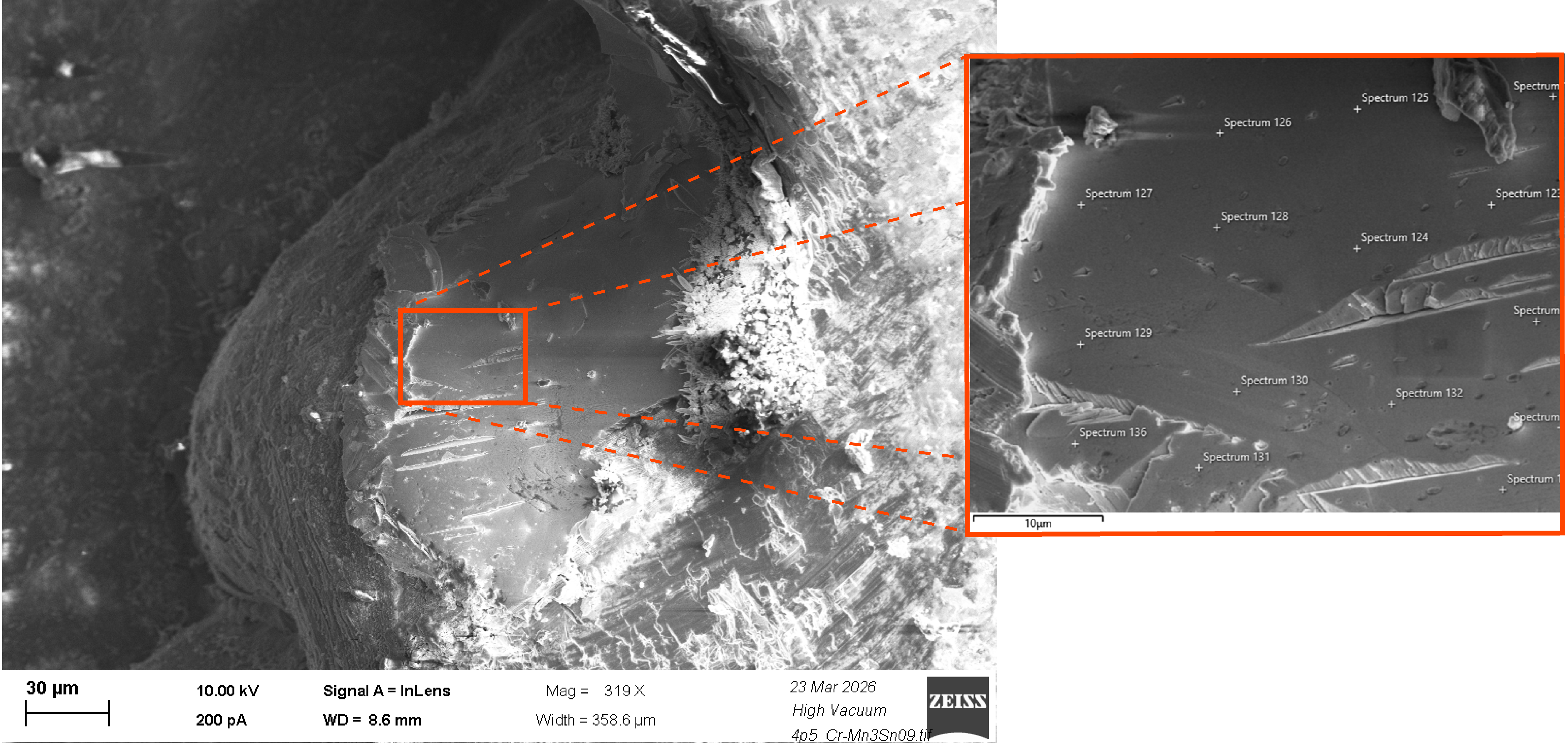}
    \caption{SEM-EDX showing a Bi inclusion found on the $x=0.05$ crystal. No alloying inside the crystal was found, see text.}
    \label{fig:Bi_inclusion}
\end{figure}

\begin{table}[H]
\centering
\normalsize
\caption{Mean atomic percentages (at\%) of the elements in the inclusion at the bottom notch of the crystal, determined by SEM-EDS. Values are means over $N$ point spectra acquired across the inclusion, each normalised over the four fitted elements (Cr, Mn, Sn, Bi); no light elements were included in the fit. Uncertainties are $2\sigma$ standard errors of the mean across the point spectra and therefore describe the compositional spread over the inclusion rather than the counting statistics of any single spectrum. The map sum spectrum of the same region is excluded from the mean, as it is not independent of the individual points.}
\label{tab:bi_inclusion_eds}

\setlength{\tabcolsep}{4pt}
\begin{tabular}{|c|c|c|c|c|c|}
\hline
\makecell[l]{Crystal E, $x$=0.05} &
\makecell[l]{$N$} &
\makecell[l]{Cr\\(at\%)} &
\makecell[l]{Mn\\(at\%)} &
\makecell[l]{Sn\\(at\%)} &
\makecell[l]{Bi\\(at\%)} \\
\hline\hline
Marked region Figure S2 & 15 & $1.58 \pm 1.31$ & $2.21 \pm 1.23$ & $0.80 \pm 0.66$ & $95.41 \pm 1.91$ \\ \hline
\end{tabular}
\end{table}


\newpage

\section*{Supplementary Note 2: magnetometry}
\noindent Figure S3 reproduces Figure 2 from the main text, now including crystals C and D. As expected from the ICP-OES data (see Supplementary Note 1), their behaviour is very similar to that of crystal E, stabilising the AFM-IT phase.

\begin{figure}[H]
    \centering
    \includegraphics[width=0.5\linewidth]{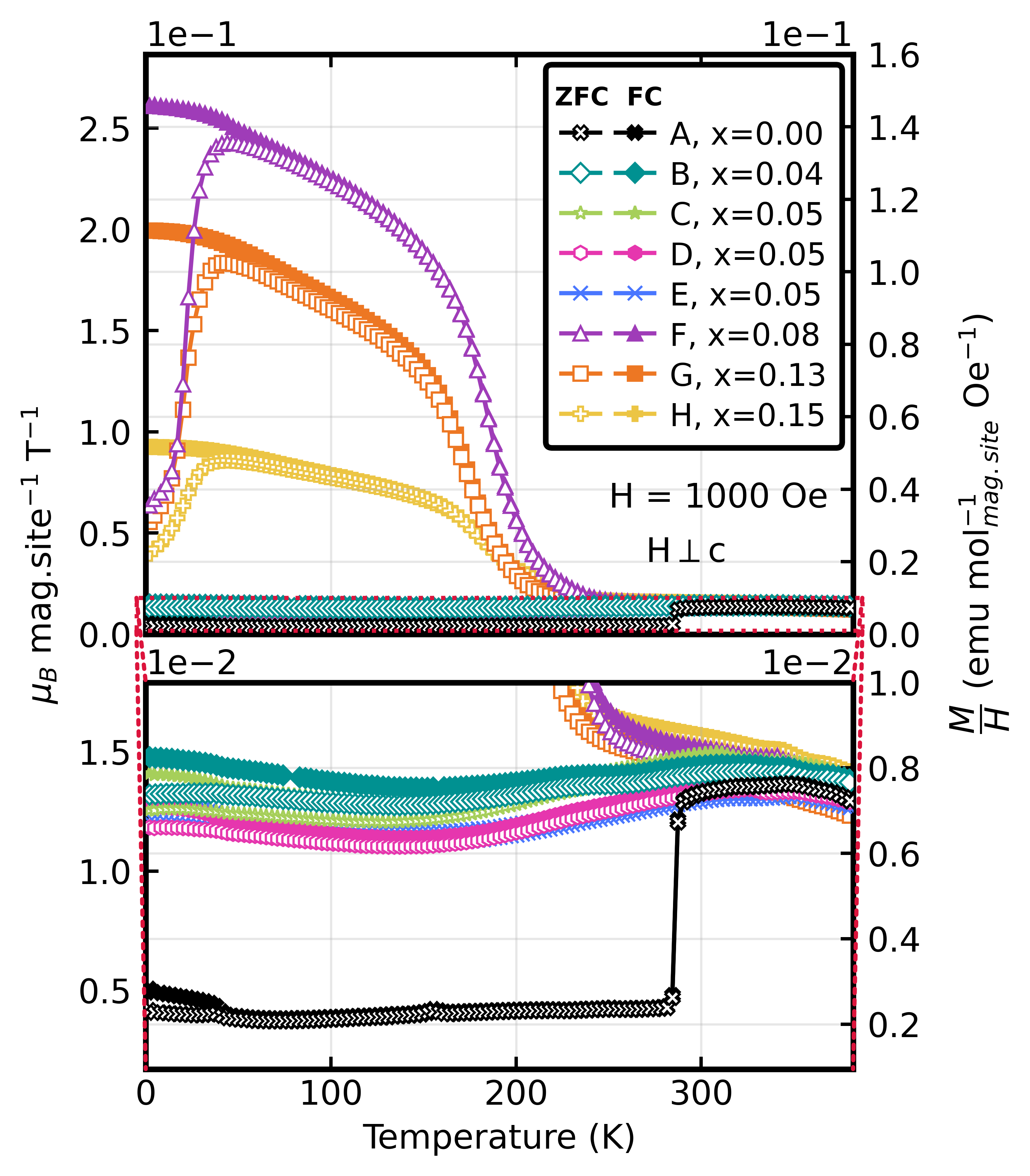}
    \caption{Magnetic susceptibility data for all crystals including crystals C and D that were omitted from main text.}
    \label{fig:SI_susc}
\end{figure}

\clearpage

\begin{figure}[H]
    \centering
    \includegraphics[width=0.4\linewidth]{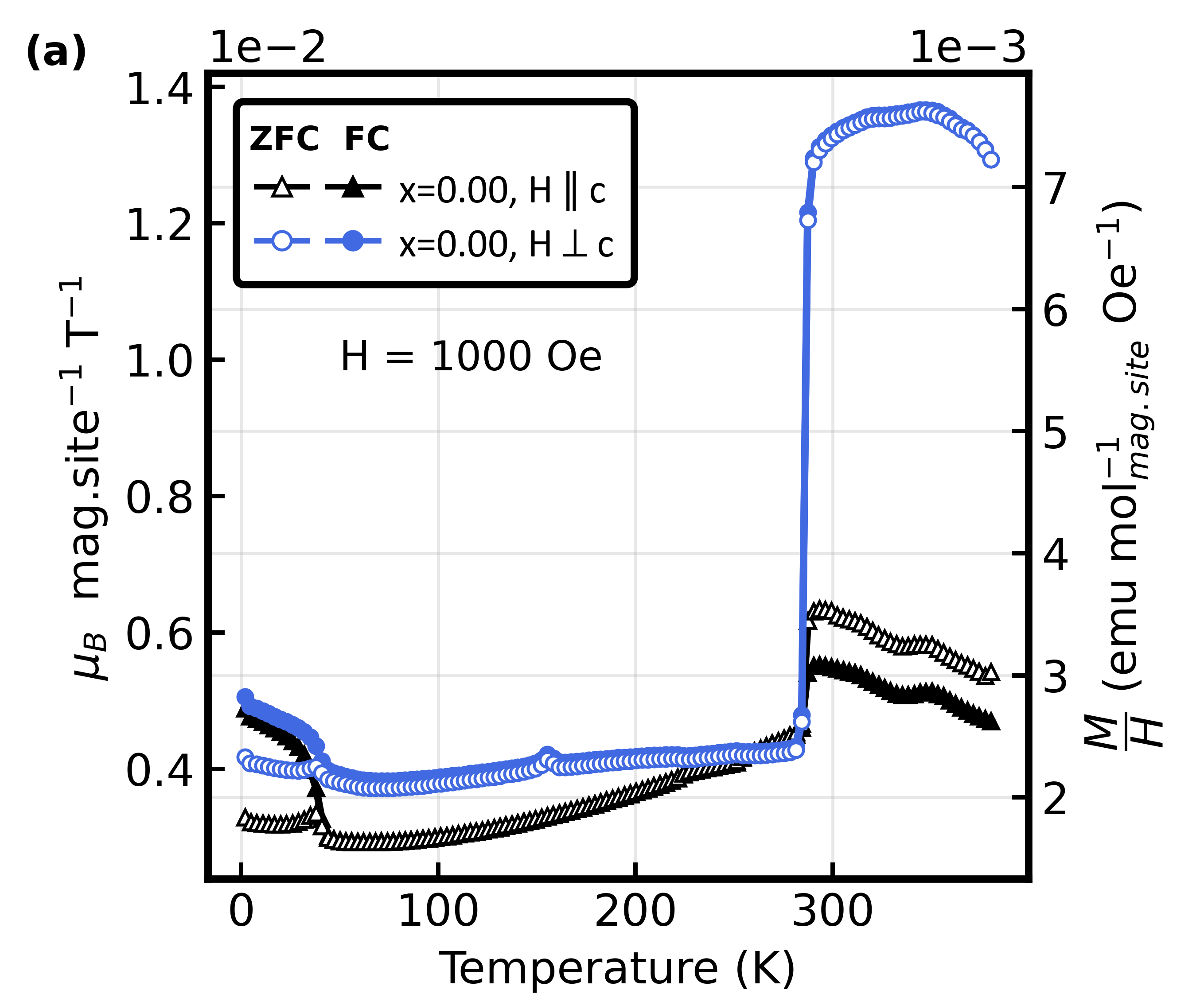}
    \includegraphics[width=0.4\linewidth]{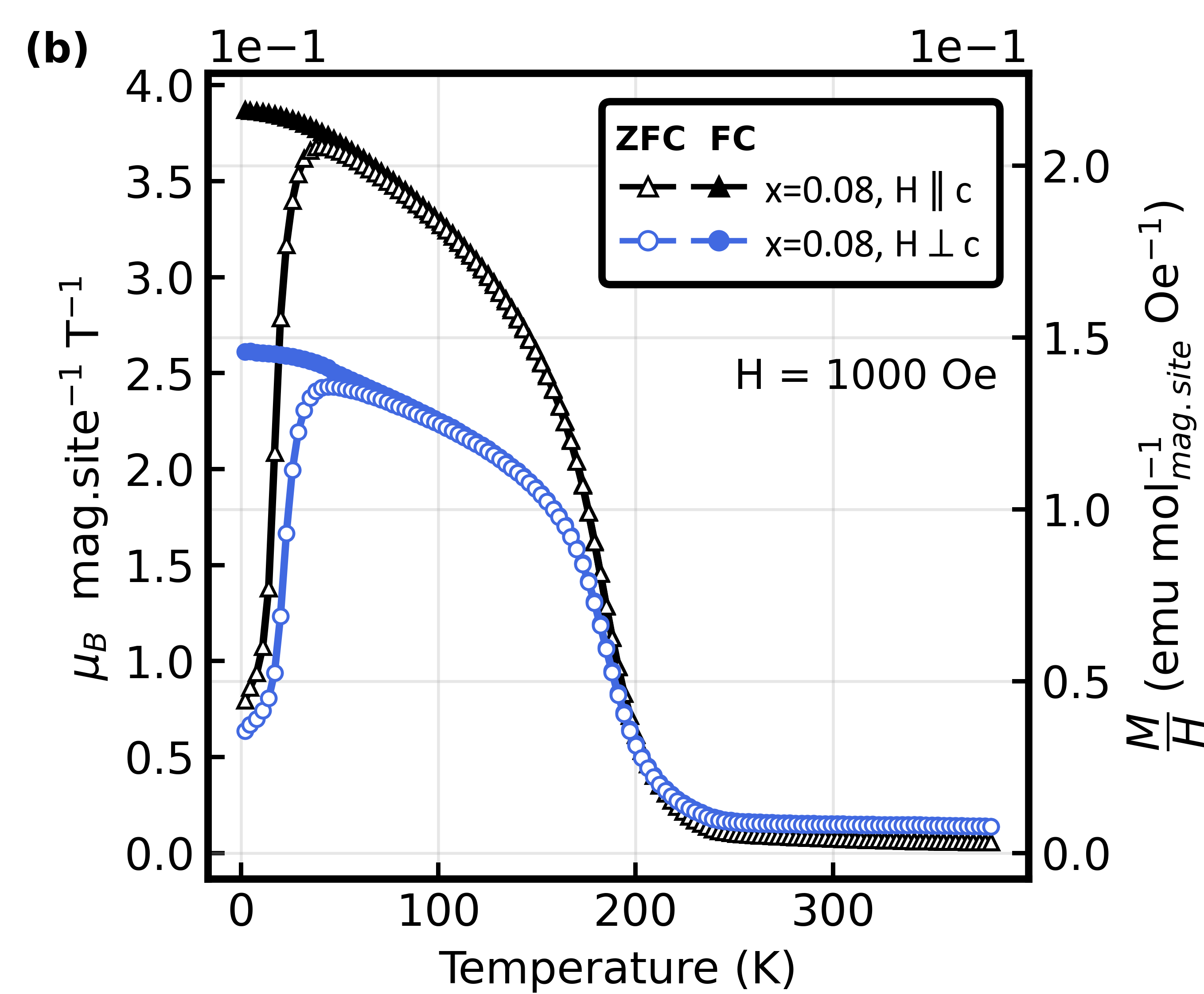}
    \caption{ZFC and FC magnetic susceptibility measurements for two crystal orientations in crystals with doping (a) $x=0.00$ and (b) $x=0.08$. }
    \label{fig:para_perp_susc}
\end{figure}

\begin{figure}[H]
    \centering
    \includegraphics[width=0.98\textwidth]{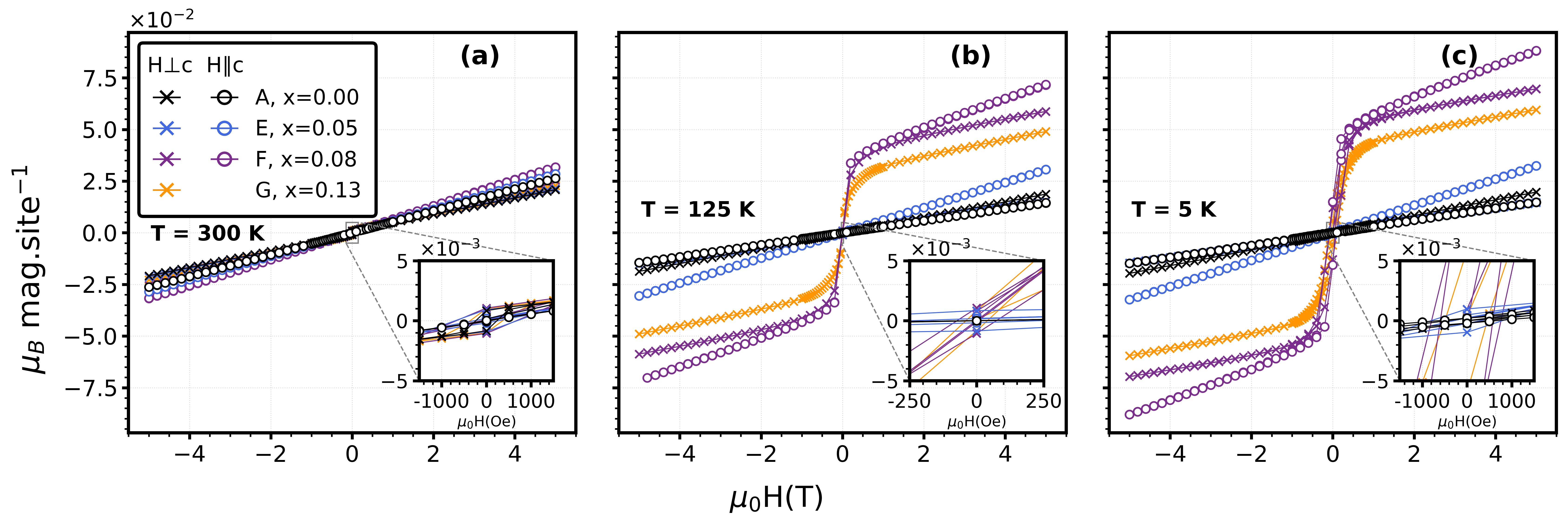}
    \caption{Magnetisation as function of applied magnetic field for two crystal orientations in crystals with doping $x=0.00$, $x=0.05$ (crystal E), $x=0.08$ and $x=0.13$ at (a) 300 K, (b) 125 K and (c) 5 K. Insets show zoom in at origin.}
    \label{fig:para_perp_MvsH}
\end{figure}

The susceptibility measurements of Figure 2 from the main text are all measured with H$\perp$c. On probing both susceptibility and magnetisation sweeps in both H$\perp$c and H$\parallel$c orientations, shown in Figures \ref{fig:para_perp_susc} and \ref{fig:para_perp_MvsH}, we see that the anisotropy between the orientations changes at $x = 0.08$. Tracking the slope of the magnetisation at high fields, the parent compound is, like $x = 0.08$, easier magnetised with H$\parallel$c at RT. However, on reducing temperature this trend reverses in the undoped parent only, indicating that doping promotes moment orientation along the $c$-axis.

\clearpage

\section*{Supplementary Note 3: anomalous Hall measurements}

\begin{figure}[H]
\centering
\includegraphics[width=0.7\textwidth]{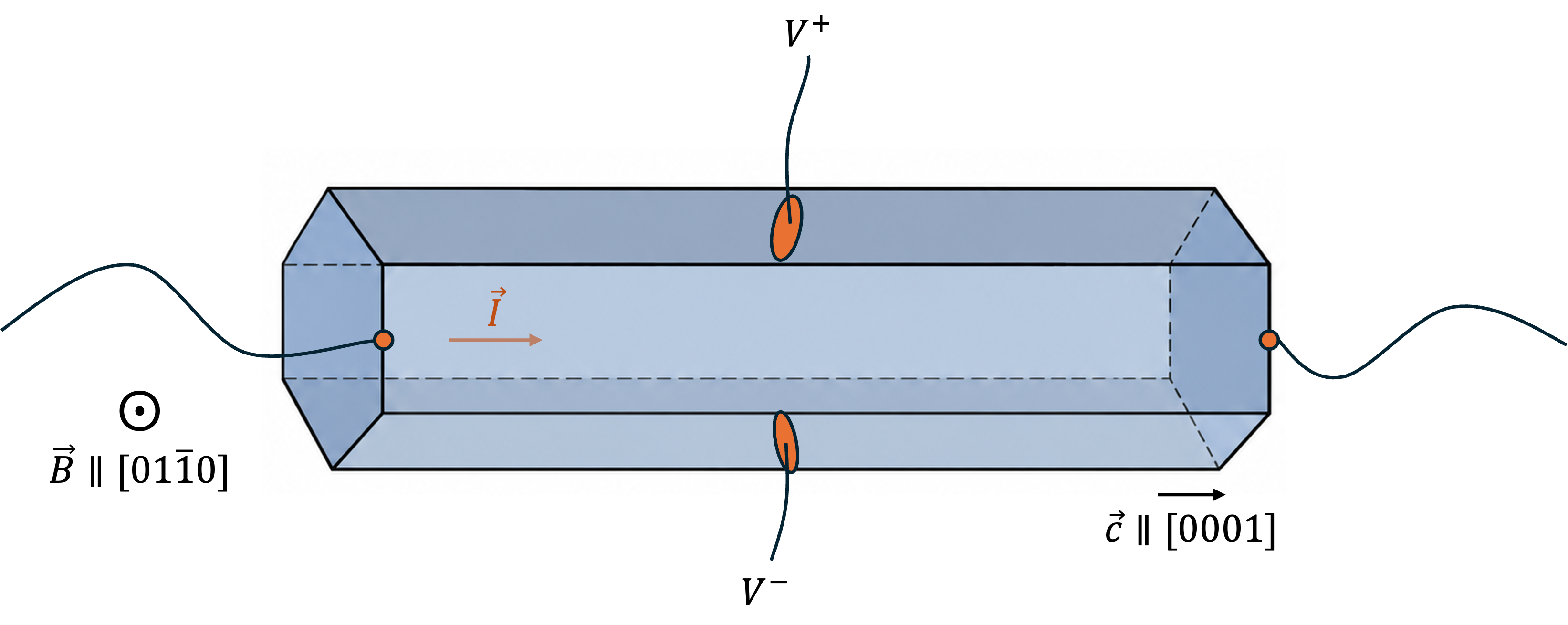}
\caption{Electrical configuration of van der Pauw geometry.}
\label{fig:vdP}
\end{figure}

\noindent Anomalous Hall transport measurements were performed in the van der Pauw geometry (see Figure \ref{fig:vdP}). The bulk crystals were electrically contacted with 4 platinum wires, which were glued with silver conductive epoxy from RS component\textsuperscript{\textregistered}. To perform Hall measurements, we pass current along the $c$-axis, with two transverse probes as V\textsuperscript{+} and V\textsuperscript{-}.

Table \ref{tab:CrMn3Sn_AHE_switching} summarizes the coercive field ($B_{c}$) and anomalous Hall amplitude of different crystals at room temperature ($T=300$ K) from Figure 3(b) in the main text.

\begin{table}[H]
\centering
\normalsize
\caption{Table summarising the Anomalous Hall effect (AHE) switching fields and resistivity amplitude of doped \CrMnSn\ single crystals, measured at room temperature ($T=300$ K), following Figure 3 of the main text.}
\label{tab:CrMn3Sn_AHE_switching}
\setlength{\tabcolsep}{4pt}
\begin{tabular}{|c|c|c|}
\hline
\makecell[c]{Doping} &
\makecell[c]{Coercive field $B_{\rm c}$ (Oe)} &
\makecell[c]{$\rho_{H}$ amplitude (\text{$\mu\Omega\cdot$ cm})}\\
\hline\hline
$x=0.04$ & $\sim 84$ & 3.34\\
\hline
$x=0.05$ & $\sim 20$ & 2.34\\
\hline
$x=0.08$ & $\sim 504$ & 2.66\\
\hline
$x=0.13$ & $\sim 434$ & 2.60\\
\hline
\end{tabular}
\end{table}

\clearpage

\section*{Supplementary Note 4: Ordinary Hall component of \NoCaseChange{\ce{(Cr_{0.13}Mn_{0.87})_{3}Sn}}}\label{sec:OHE}

\begin{figure}[H]
    \centering
    \includegraphics[width=0.7\linewidth]{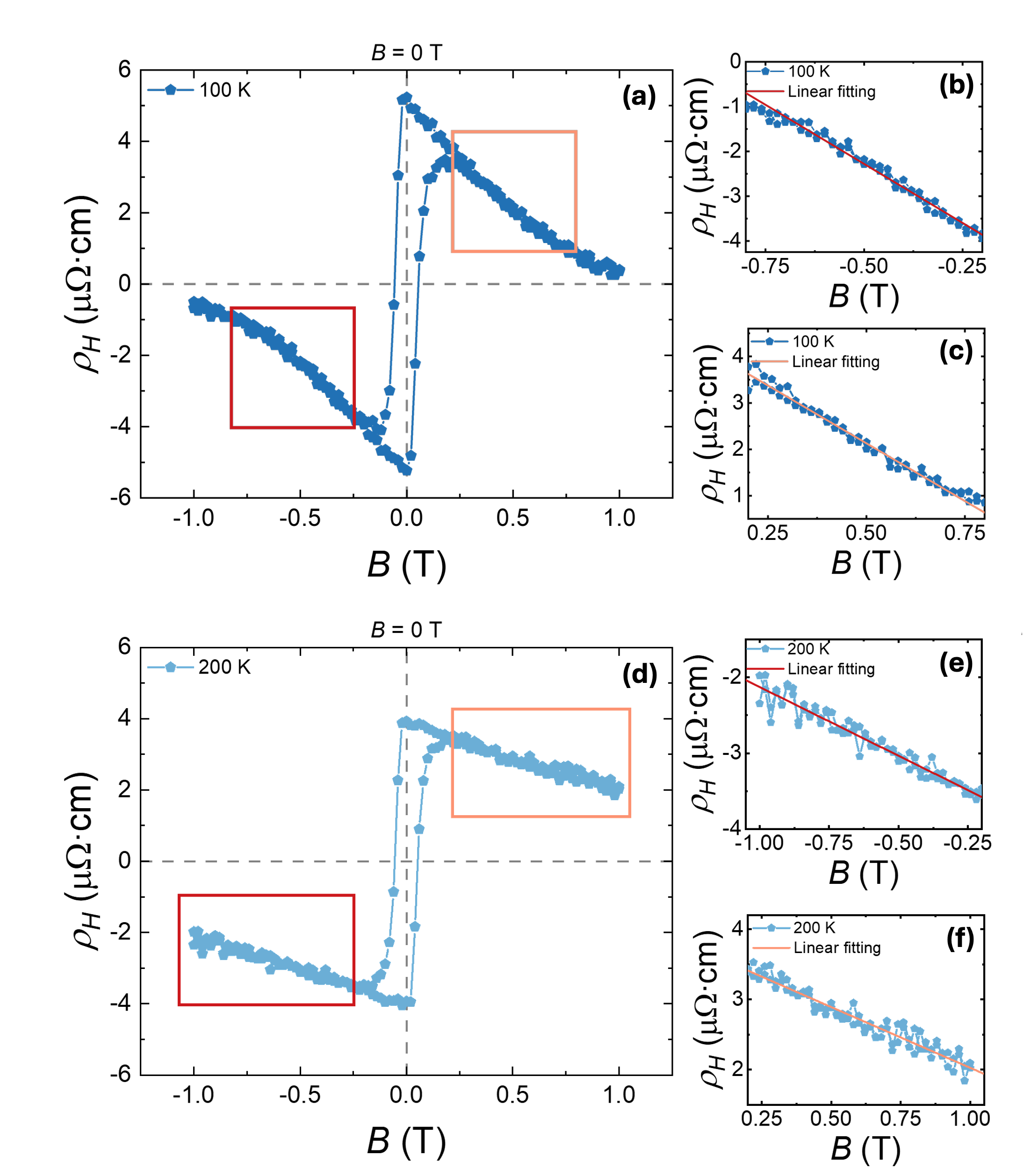}
    \caption{Anomalous Hall measurements for \CrMnSnEight\ at (a) $T=100$ K and (d) $T=200$ K. (b) and (c) show linear fitting of lower branch (red box in (a)) and upper branch (orange box in (a)) respectively. (e) and (f) show linear fitting of lower branch (red box in (d)) and upper branch (orange box in (d)) respectively. }
    \label{fig:CarrierDensity}
\end{figure}

\noindent Figure \ref{fig:CarrierDensity}(a) and (d) show the Hall measurements performed on the $x=0.13$ crystal at 100 K and 200 K respectively. For antiferromagnets, the Hall response ($\rho_{H}$) is contributed by two parts: the ordinary Hall effect ($\rho_{H}^{OHE}(B)$) and the anomalous Hall effect ($\rho_{H}^{AHE}(B)$), where
\begin{equation}\label{eqn:OHE}
    \rho_{H}^{OHE}(B)\approx R_{0}\cdot B
\end{equation}
in a single-band approximation. Hence, away from the anomalous Hall regime, the Hall effect is expected to show a linear response to an external magnetic field. Figure \ref{fig:CarrierDensity}(a) and (d) show the linear Hall response as a function of an external magnetic field (highlighted with red and orange boxes). By extracting the slope of the linear parts, $R_{0}$, of the Hall response, we can then calculate the carrier density by:
\begin{equation}\label{eqn:carrierdensity}
    n = \frac{1}{eR_{H}}
\end{equation}
where $e=1.602\times10^{-19}$ C. From equation \ref{eqn:carrierdensity}, the carrier density of the $x=0.13$ crystal can be summarized in Table \ref{tab:carrierdensity}.

\begin{table}[H]
\normalsize
\caption{Carrier density of \ce{(Cr_{0.13}Mn_{1-0.13})_{3}Sn} at 100 K and 200 K.}
\label{tab:carrierdensity}
\centering

\setlength{\tabcolsep}{4pt}
\begin{tabular}{|c|c|c|}
\hline
\makecell[l]{Temperature (K)} &
\makecell[l]{$n$ ($\times10^{20}$\ce{cm^{-3}})} &
\makecell[l]{Charge carrier type}\\
\hline\hline
100 & 1.218 & Hole \\ \hline
200 & 3.528 & Hole \\ \hline
\end{tabular}
\end{table}

It is worth pointing out that the carrier density in the $x=0.13$ crystal is about two orders of magnitude lower than that of the Mn-rich \ce{Mn_{3+\delta}Sn} crystal from Nakatsuji \textit{et al.}\cite{Nakatsuji2015}. Meanwhile, there is a clear sign of non-linearity of the Hall response at 100 K in higher magnetic field, which implies a breakdown of the single-band approximation from equation \ref{eqn:OHE}. Therefore, it is reasonable to assume that the Fermi surface of the $x=0.13$ crystal is closer to the Dirac point, such that multiple hole and electron pockets are contributing to the charge carriers. Further investigations on the band structures of this crystal, such as angle-resolved photoemission spectroscopy (ARPES) and density function theory (DFT), can help clarify this hypothesis.

\section*{Supplementary References}
\bibliographystyle{apsrev4-1}
\bibliography{bibliography}